\documentclass[aps,prl,showpacs,groupedaddress,floatfix,nofootinbib]{revtex4}

\usepackage{graphicx}
\usepackage{dcolumn}

\begin{document}

\title{On recent calculations of resonances for the Stark effect in hydrogen}
\author{Francisco M. Fern\'andez}\email{fernande@quimica.unlp.edu.ar}

\author{Javier Garcia}

\affiliation{INIFTA (UNLP, CCT La Plata-CONICET), Blvd. 113 y 64 S/N, \\
Sucursal 4, Casilla de Correo 16, 1900 La Plata, Argentina}

\begin{abstract}
We show that the resonances of the Stark effect in hydrogen
reported in Phys. Rev. A \textbf{88}, 022509 (2013) are
considerably less accurate than the number of digits appear to
suggest. In particular, the imaginary part of the lowest resonance
is several orders of magnitude greater than it should be. We
compare the results of that paper with those provided by the
Riccati-Pad\'{e} method, perturbation theory and an asymptotic
expansion for the resonance width. The inaccuracy of those results
can be traced back to the lack of precision in the calculation of
the matrix elements of the secular equation. We carry out a more
accurate calculation with the same method and show that the
agreement with earlier results of other authors is greatly
improved.
\end{abstract}

\pacs{31.15.ac, 32.60.+i, 52.70.Ds}

\maketitle

In a recent paper Fern\'{a}ndez-Menchero and Summers\cite{FS13}
obtained the complex eigenvalues and eigenfunctions of the
Hamiltonian operator for the hydrogen atom in a uniform electric
field. They resorted to the Laguerre-mesh basis set proposed by
Lin and Ho\cite{LH11} for the treatment of the Yukawa potential in
a uniform electric field and the complex-rotation
method\cite{CHHRSW78}. For brevity we will refer to this method as
CRLM from now on. They compared their results with those obtained
by Lin and Ho\cite {LH11}, Kolosov\cite{K87}, Rao and
Li\cite{RL95} and Ivanov\cite{I01} and overlooked the earlier
impressive calculations of Benassi and Grecchi\cite {BG80} and the
accurate results obtained by Fern\'{a}ndez\cite{F96b}. Benassi and
Grecchi resorted to a basis set of confluent hypergeometric
functions that is suitable when the Schr\"{o}dinger equation is
written in squared parabolic coordinates. On the other hand,
Fern\'{a}ndez applied the Riccati-Pad\'{e} method (RPM) that does
not require the use of complex coordinates.

The results of Fern\'{a}ndez-Menchero and Summers\cite{FS13} for
the excited states deviate somewhat from the other
ones\cite{LH11,K87,RL95,I01} that appear to be in closer agreement
among them. On the other hand, they were unable to compare their
estimate of the lowest resonance for the field strength
$F=0.005\,a.u$ because there does not appear to be any independent
calculation available.

Fern\'{a}ndez-Menchero and Summers\cite{FS13} stressed that
perturbation theory (PT) is unsound because the Stark effect does
not fulfill the conditions for the application of such an
approach. They are obviously unaware of the fact that the
straightforward perturbation series is asymptotic to the real part
of the resonances which in some cases can be obtained with
reasonable accuracy by judicious truncation to a suitable number
of terms. Besides, Adams\cite{A94} calculated the real and
imaginary parts of the lowest resonance by means of analytic
continuation and Pad\'{e} summation and Jentschura\cite{J01}
resorted to Borel-Pad\'e summation for obtaining several selected
resonances.

The purpose of this comment is to show that the resonances
reported by those authors are considerably less accurate than the
number of digits appear to indicate while the width for the lowest
resonance is several orders of magnitude greater than it should
be. The inaccuracy of their results, which accounts for the
discrepancy with respect to those of other authors mentioned
above, can be traced back to the lack of precision in the
calculation of the matrix elements in the secular equation. To
verify this conjecture we carry out a more accurate calculation
with the CRLM. In addition to it, we also resort to the RPM, PT
and the asymptotic formula for the width of the lowest resonance
derived by Benassi and Grecchi\cite{BG80}.

The Schr\"{o}dinger equation in atomic units is
\begin{eqnarray}
H\psi &=&E\psi  \nonumber \\
H &=&\frac{1}{2}\nabla ^{2}+\frac{1}{r}-Fz,  \label{eq:Schro_Stark}
\end{eqnarray}
where $F$ is the intensity of the uniform electric field assumed to be
directed along the $z$ axis.

This equation is separable in parabolic coordinates
\begin{eqnarray}
x &=&\sqrt{\xi \eta }\cos \phi ,\;y=\sqrt{\xi \eta }\sin \phi ,\;z=\frac{\xi
-\eta }{2}  \nonumber \\
\xi  &\geq &0,\;\eta \geq 0,\;0\leq \phi \leq 2\pi ,  \label{eq:parabolic}
\end{eqnarray}
but the authors decided to treat the Schr\"{o}dinger equation for the
resulting Hamiltonian
\begin{equation}
H=-\frac{2}{\xi +\eta }\left[ \frac{\partial }{\partial \xi }\left( \xi
\frac{\partial }{\partial \xi }\right) +\frac{\partial }{\partial \eta }%
\left( \eta \frac{\partial }{\partial \eta }\right) \right] -\frac{1}{2\,\xi
\eta }\frac{\partial ^{2}}{\partial \phi ^{2}}-\frac{2}{\xi +\eta }+F\frac{%
\xi -\eta }{2},  \label{eq:H_parabolic}
\end{equation}
as nonseparable. To this end they proposed the variational ansatz
\begin{eqnarray}
\psi \left( \xi ,\eta ,\phi \right)  &=&\frac{1}{\sqrt{2\pi }}e^{im\phi
}\sum_{k=1}^{N}\sum_{l=1}^{N}c_{klm}e^{-\frac{\xi +\eta }{2}}\left( \xi \eta
\right) ^{\frac{|m|}{2}}\Lambda _{Nk}(\xi )\Lambda _{Nl}(\eta )
\label{eq:ansatz_FS} \\
\Lambda _{Nk}(x) &=&(-1)^{k}\sqrt{x_{k}}\frac{L_{N}(x)}{x-x_{k}},
\end{eqnarray}
where $L_{N}(x)$ is the Laguerre polynomial of degree $N$ and
$x_{k}$ its $k$th zero. In order to obtain the resonances they
resorted to the well known complex rotation method\cite{CHHRSW78}
that in this case is given by the particular complex-scaling
transformation $(\xi ,\eta )\rightarrow (e^{i\vartheta }\xi
,e^{i\vartheta }\eta )$. The eigenvalues and expansion
coefficients are given by the secular equation
\begin{equation}
(\mathbf{H}-E\mathbf{S})\mathbf{C}=0,  \label{eq:secular}
\end{equation}
where the elements of the $N^{2}\times N^{2}$ matrices
$\mathbf{H}$ and $\mathbf{S}$ are explicitly shown
elsewhere\cite{FS13} and the elements of the column vector
$\mathbf{C}$ are the coefficients $c_{klm}$. Note that the
integrals appearing in the matrix elements of both $\mathbf{H}$
and $\mathbf{S}$ should be calculated numerically and when we
increase $N$ we have to calculate all those integrals again.

On the other hand, in order to apply the RPM we write
\begin{equation}
\psi (\xi ,\eta ,\phi )=(\xi \eta )^{-1/2}u(\xi )v(\eta )e^{im\phi
},\;m=0,\pm 1,\pm 2,\ldots ,  \label{eq:psi_parabolic}
\end{equation}
and obtain two equations of the form
\begin{equation}
\left( \frac{d^{2}}{dx^{2}}+\frac{1-m^{2}}{4x^{2}}+\frac{E}{2}-\sigma \frac{F%
}{4}x+\frac{A_{\sigma }}{x}\right) \Phi (x)=0,
\label{eq:separated_parabolic}
\end{equation}
where $\sigma =\pm 1$ and $A_{+}=A$ and $A_{-}=1-A$ are separation
constants. When $\sigma =1$, $x=\xi $ and $\Phi (\xi )=u(\xi )$;
when $\sigma =-1$, $x=\eta $ and $\Phi (\eta )=v(\eta )$.

The regularized logarithmic derivative
\begin{equation}
f(x)=\frac{s}{x}-\frac{\Phi ^{\prime }(x)}{\Phi (x)},\;s=\frac{|m|+1}{2},
\label{eq:f(x)}
\end{equation}
can be expanded in a Taylor series
\begin{equation}
f(x)=\sum_{j=0}^{\infty }f_{j}x^{j},  \label{eq:f(x)_exp}
\end{equation}
where the coefficients $f_{j}$ are polynomial functions of $E$ and
$A$. The details of the method are outlined elsewhere\cite{F96b};
here it suffices to say that we construct the Hankel determinant
\begin{equation}
H_{D}^{d}(E,A,F)=\left|
\begin{array}{cccc}
f_{d+1} & f_{d+2} & \ldots  & f_{D+d} \\
f_{d+2} & f_{d+3} & \ldots  & f_{D+d+1} \\
&  & \ddots  &  \\
f_{D+d} & f_{D+d+1} & \ldots  & f_{2D+d-1}
\end{array}
\right| ,  \label{eq:Hankel}
\end{equation}
where $D=2,3,\ldots $ and $d=0,1,\ldots $ is kept fixed. The approximate
eigenvalues are given by the set of nonlinear equations
\begin{equation}
H_{D}^{d}(E,A,F)=H_{D}^{d}(E,A-1,F)=0.  \label{eq:RPM_quantization}
\end{equation}

The main advantage of the RPM is the enormous rate of convergence of the
approximate eigenvalues $E^{[D,d]}$ which enables us to obtain very accurate
results with determinants of relatively small dimension. However, the great
number of roots in the neighborhood of each eigenvalue makes it difficult to
find the optimal sequence that converges to it. We can mention two other
disadvantages of the RPM: first, it only applies to separable problems;
second, it does not give the eigenfunction but its logarithmic derivative.
Consequently, it is not practical for the calculation of physical properties
that are defined in terms of the eigenfunctions. In spite of these
disadvantages the RPM is an extremely useful tool to produce accurate
results for benchmark purposes in those cases where it can be applied. The
Stark effect in hydrogen is one such problem as shown in what follows.

\begin{figure}[H]
\begin{center}
\includegraphics[width=9cm]{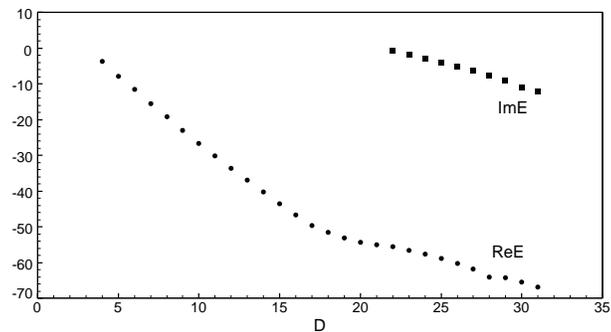}
\end{center}
\caption{Convergence of the RPM for the lowest resonance when $F=0.005$}
\label{fig:F0005}
\end{figure}

In order to test the rate of convergence we calculate $\log \left|
\left( \alpha ^{[D+1,d]}-\alpha ^{[D,d]}\right) /\alpha
^{[D+1,d]}\right| $ for $\alpha =\mathrm{Re}E$ or $\alpha
=\mathrm{Im}E$. Here we restrict ourselves to $d=0$. The results
for the lowest resonance when $F=0.005$ shown in
Fig.~\ref{fig:F0005} strongly suggest that present RPM calculation
may be remarkably accurate.

The converged RPM resonance, truncated to a reasonable number of
digits, is shown in Table~\ref{tab:F0005} together with the CRLM
one\cite{FS13}. The discrepancy between the real parts (1 in
$10^6$) may not appear to be so serious at first sight, but the
imaginary parts differ in many orders of magnitude predicting
considerably different lifetimes for the same metastable state.
Our estimated value of $\mathrm{Im}E$ is in perfect agreement with
the analytic asymptotic formula derived by Benassi and
Grecchi\cite{BG80}:
\begin{equation}
\mathrm{Im}E\sim -2F^{-1}e^{-2/(3F)}\left(
1-8.916F+25.57F^{2}+O(F^{3})\right) ,  \label{eq:ImE_asymp}
\end{equation}
also shown in Table~\ref{tab:F0005}. From a practical point of
view this state may be considered stable because there is no
experimental way of determining such a long lifetime. However, it
is worth stressing the fact that the RPM enables us to calculate
it.

In principle we expect that a properly truncated perturbation
series will exhibit an accuracy of the order of $\left|
\mathrm{Im}E\right|$. On summing the first 130 terms of the
perturbation series calculated by means of the hypervirial
perturbative method\cite{F00} we obtained the following result:

\begin{eqnarray}
E^{PT}&=&-0.5000562847937929693317739476914328819632509273188913726,
\nonumber \\
\mathrm{Re}E^{RPM}&=&-0.50005628479379296933177394769143288196325092731889137262135731287257,
\end{eqnarray}
that agrees with the RPM one to the last digit. Obviously, this
result is more accurate than the CRLM one and most probably than
any other calculation based on diagonalization and complex
rotation.

In our opinion it is quite difficult to obtain so small imaginary
parts of resonances by means of standard methods based on
diagonalization and complex rotation which may probably be the
reason why it was not included in the earlier calculations chosen
for comparison\cite{LH11,K87,RL95,I01}. This fact clearly shows
that the RPM is a most valuable tool for testing other
calculations on separable problems.

The discrepancy between the results of Fern\'andez-Menchero and
Summers and those chosen for comparison is probably due to
insufficient accuracy in the calculation of the CRLM integrals. In
order to test this conjecture we calculated those integrals with
greater accuracy (15 digits) by means of a computer algebra system
that enables, in principle, unlimited precision and then solved
the CRLM secular equation in the usual way for $N=30$. Present
CRLM lowest resonance, also shown in Table~\ref{tab:F0005}, is in
perfect agreement with the RPM result. Table~\ref{tab:F0005b}
shows that present CRLM resonances are in better agreement with
those of the other authors\cite{LH11,K87,RL95,I01} (for brevity we
only show the results of Lin and Ho\cite{LH11}). In passing we
mention that the width of the state $\left| {}\right.
2\;1\;0\left. {}\right\rangle$ calculated by the other
authors\cite{LH11,K87,RL95,I01} was incorrectly transcribed by
Fern\'andez-Menchero and Summers.

Typically the minimum value of $\left|\mathrm{Im}E\right|$ that
one can calculate by a diagonalization method is of the order of
the accuracy of the eigenvalue. The accuracy of the results of
Fern\'andez-Menchero and Summers is roughly of the order of
$10^{-6}$ which is the reason why they estimated such width for
the lowest resonance. If we take into account that other physical
effects (magnetic-field perturbation, fine structure, etc.)
already mentioned by the authors are of the order of $10^{-4}$ we
conclude that the lack of precision just mentioned may probably
not be serious for most practical applications. However, we have
clearly shown that the CRLM results can be considerably improved
by increasing the accuracy of the integrals that provide the
elements of the matrices in the secular equation.

We add that PT and the asymptotic formula\cite{BG80} are
remarkably useful for the calculation of extremely sharp
resonances. Unfortunately, while we have the perturbation series
for all the resonances\cite{F00} the asymptotic formula is only
available for the lowest one\cite{BG80}. It should be mentioned
that, even for the small field intensity considered here, PT is
not so accurate for higher resonances unless one resorts to a
suitable summation method\cite{A94,J01}. On the other hand, the
RPM yields eigenvalues of similar accuracy to the one shown here.

\begin{table}[H]
\caption{Lowest resonance for $F=0.005$}
\label{tab:F0005}
\begin{center}
\begin{tabular}{lll}
\hline & \multicolumn{1}{c}{Re$E$} & \multicolumn{1}{c}{$\Gamma$} \\
\hline

CRLM \cite{FS13} & $-0.5000553416$ & $0.8944475605\times
10^{-7}$ \\

Present CRLM & $-0.500056284793$ & $< 1\times 10^{-13}$ \\

\multicolumn{1}{l}{RPM} & $-0.5000562847938$ &
$9.49802741674\times 10^{-56}$ \\

\multicolumn{1}{l}{Asymptotic} &  & $9.4983\times 10^{-56}$ \\

\end{tabular}
\end{center}
\end{table}

\begin{table}[H]
\caption{Resonances for $F=0.005$} %
\label{tab:F0005b}
\begin{center}
\begin{tabular}{lll}
\hline & \multicolumn{1}{c}{Re$E$} & \multicolumn{1}{c}{$\Gamma$} \\
\hline

\multicolumn{1}{l}{$\left| {}\right. 2\ -1\ 0\left.
{}\right\rangle$} & & \\
CRLM \cite{FS13} & $-0.1426203564$ & $1.057292433\times 10^{-4}$
\\

Present CRLM & $-0.1426186075727079$ & $1.05944463673\times 10^{-4}$ \\

Ref. \cite{LH11} & $-0.1426186076$ & $1.059444711\times 10^{-4}$ \\

\multicolumn{1}{l}{$\left| {}\right. 2\ 1\ 0\left. {}\right\rangle
$} & & \\

CRLM \cite{FS13}  & $-0.1120633027$ & $4.930560122\times 10^{-6}$
\\

Present CRLM & $-0.1120619240019936$ &
$5.72936843930\times 10^{-6}$                      \\
Ref. \cite{LH11} & $-0.1120619240$ & $5.72939466\times 10^{-6}$ \\

\multicolumn{1}{l}{$\left| {}\right. 2\ 0\ 1\left. {}\right\rangle
$} &  & \\

CRLM \cite{FS13} & $-0.1271464039$  & $2.671348551\times 10^{-5}$
\\

Present CLRM & $-0.127146612703972$ & $2.6152854466430\times
10^{-5}$ \\

Ref. \cite{K87}  & $-0.127 146 612$ & $2.61528545\times 10^{-5}$

\end{tabular}
\end{center}
\end{table}

\end{document}